\documentstyle[emulateapj,psfig]{article}
\lefthead{Levenson et al.}
\righthead{The PDR M16}

\def\h2{H$_2$}
\def\kms{\ifmmode{\rm \,km\,s^{-1}}\else${\rm \,km\,s^{-1}}$\fi}
\def\pcc{\ifmmode{\rm \,cm^{-3}}\else${\rm \,cm^{-3}}$\fi}

\def\deg{\ifmmode^\circ\else$^\circ$\fi}

\def\gs{{_>\atop^{\sim}}}
\def\micron{\ifmmode{\mu{\rm m}}\else~$\mu$m\fi}

\def\arcsec{\ifmmode {^{\prime\prime}}\else $^{\prime\prime}$\fi}
\def\arcmin{\ifmmode {^\prime}\else $^\prime$\fi}
\def\han {\mbox{{\rm H}$\alpha$}}
\def\ha{\han}
\def\brg {{\rm Br}$\gamma$}

\begin{document}
\submitted{To appear in ApJ Letters, April 10, 2000}
\title{
Hot Stars and Cool Clouds: 
The Photodissociation Region M16$^1$ 
}  

\author{ N. A. Levenson\altaffilmark{2}, 
James R. Graham\altaffilmark{3},
Ian S. McLean\altaffilmark{4}, 
E. E. Becklin\altaffilmark{4}, 
Donald F. Figer\altaffilmark{5}, 
Andrea M. Gilbert\altaffilmark{3}, 
James E. Larkin\altaffilmark{4}, 
Harry I. Teplitz\altaffilmark{6,}\altaffilmark{7}, 
Mavourneen K. Wilcox\altaffilmark{4}}
\altaffiltext{1}{
Data presented herein were obtained at the W.M. Keck Observatory, 
which is operated
as a scientific partnership among the California Institute of Technology, 
the University of California and the National Aeronautics and Space 
Administration.  The Observatory was made possible by the
generous financial support of the W.M. Keck Foundation.
}
\altaffiltext{2}{Department of Physics and Astronomy, Johns Hopkins University,
                 Baltimore, MD  21218}
\altaffiltext{3}{Department of Astronomy,  
University of California, Berkeley, CA, 94720-3411}
\altaffiltext{4}{Department of Physics and Astronomy, 
University of California, Los Angeles, CA, 90095-1562 }
\altaffiltext{5}{Space Telescope Science Institute, 
                  3700 San Martin Dr., Baltimore, MD 21218 }
\altaffiltext{6}{Laboratory for Astronomy and Solar Physics, Code 681, Goddard
Space Flight Center, Greenbelt MD 20771}
\altaffiltext{7}{NOAO Research Associate}

\begin{abstract}
We present high-resolution spectroscopy and images of 
a photodissociation region (PDR) in M16 obtained during commissioning of
NIRSPEC on the Keck II telescope.
PDRs play a significant role in regulating star formation, and
M16 offers the opportunity to examine the physical processes
of a PDR in detail.
We simultaneously observe both the molecular and ionized phases 
of the PDR and resolve the spatial and kinematic differences
between them.  
The most prominent regions of the PDR are viewed edge-on.
Fluorescent emission from nearby stars is the primary
excitation source, although collisions also preferentially 
populate the lowest vibrational levels of \h2.
Variations in density-sensitive emission line ratios demonstrate
that the molecular cloud is clumpy, 
with an average density $n = 3\times 10^5{\rm \,cm^{-3}}$.
We measure the kinetic temperature of the molecular region directly
and find $T_{H_2} = 930$ K.
The observed density, temperature, and UV flux imply 
a photoelectric heating efficiency of 4\%. 
In the ionized region, $n_i=5\times10^3{\rm\,cm^{-3}}$ and 
$T_{H{\sc ii}} = 9500$ K.  In the 
brightest regions of the PDR, the recombination line widths include a non-thermal component,
which we attribute to viewing geometry.
\end{abstract}


\section{Introduction}
Star formation occurs in dense regions of the interstellar
medium, and as a result, the environments of early-type stars 
often contain molecular clouds.
In addition to creating an \ion{H}{2} region, where
hydrogen is predominantly ionized,
stellar ultraviolet radiation creates 
a photodissociation region (PDR) when it interacts with molecular 
material.  The feedback between extant stars and the 
nearby clouds in PDRs is important in regulating subsequent episodes of
star formation.  The UV flux determines the ionization
fraction in molecular clouds, which in turn sets the ambipolar
diffusion rate and thereby the rate of low-mass star formation
(\cite{McK89}).  

Hollenbach \& Tielens (1997, 1999)\nocite{Hol97,Hol99}  
comprehensively review the physical and chemical processes of PDRs.
Far-ultraviolet (FUV) photons ($6 < h\nu < 13.6 {\rm\,eV}$)
dissociate molecules and excite the gas on the surface
($A_V < 3$) of the molecular cloud.  
The result is spatial stratification of molecular, atomic, and ionic
components of a given element.
At higher densities ($n\gs 10^4 {\rm\,cm^{-3}}$), collisional excitation 
also populates the excited states (\cite{Ste89}).

The columns of M16 (the Eagle Nebula) are PDRs.
Most of the mass in the nebula is in molecular clouds,
and a variety of molecular transitions are observed 
directly (\cite{Pou98}; \cite{Whi99}).  \h2\ is
observed on the cloud surfaces (\cite{All99}), and at the interface with
the \ion{H}{2}\ region, the columns are prominent in 
optical emission lines (\cite{Hes96}).
The young cluster NGC 6611 contains many massive stars,
which illuminate the columns 
with total FUV flux $16 {\rm\,erg\,s^{-1}\,cm^{-2}}$ (\cite{All99})
from a distance of about 2 pc (\cite{Hes96}) .

We simultaneously observe the spatial and
spectral relationship of the excited molecular and ionized
phases of the PDR.  
Based on these observations (described in \S2), we
determine physical conditions in \S3, 
and summarize the conclusions in \S4.

\section{Observations and Data Reduction\label{secdata}}

\begin{figure*}
\centerline{\psfig{figure=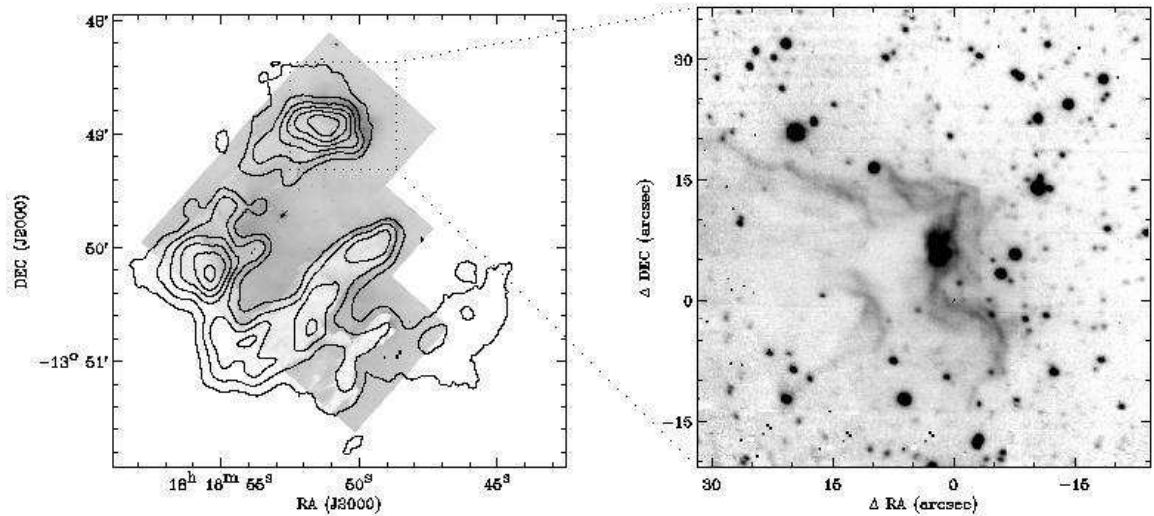,angle=-90,width=6in}}
\figcaption{(a) The {\it HST} \ha\ image of M16 (\cite{Hes96}), with contours
of integrated CO intensity ({\it solid lines}; \cite{Pou98}) and
the location of the NIRSPEC slit-viewing camera 2.12\micron\ image
({\it dotted line}) overlaid.
The \ha\ image is scaled with a square-root stretch, and
the CO contours are scaled linearly, from 50 to 300 ${\rm Jy\,km\,s^{-1}}$.
(b) Northern column of M16 at 2.12\micron\ (including \h2\ 1-0S(1)), 
observed with the NIRSPEC slit-viewing camera.  
Right ascension and declination are marked with respect to the observed
center of the spectroscopy slit, which is overlaid.  The UV flux
comes from the northwest, which corresponds to the top of the slit
in subsequent figures.  The excited molecular emission is a surface 
phenomenon, arising only at the edge of the molecular cloud.
\label{fighstcoh2}}
\end{figure*}

We obtained images and high resolution longslit spectra of the
northernmost column of M16
using NIRSPEC, the
facility near-infrared spectrometer on the Keck II 10-m telescope (\cite{McL98}).  
Figure \ref{fighstcoh2}a identifies the observed region 
on the {\it Hubble Space Telescope} 
\ha\ image of M16 (\cite{Hes96}) with contours of integrated CO intensity
(\cite{Pou98}).
We used the slit-viewing camera to image M16 on 1999 July 6 (UT), obtaining
9 exposures of 60 s each.
In addition to the \h2\ 1-0S(1) filter, 
(bandpass 2.110--2.129\micron)
the NIRSPEC-6 filter (1.558--2.315\micron) was in the optical path. 
We combined several observations of a blank field in the NIRSPEC-6 filter alone
to use as a flat field and scaled this for sky subtraction from this
crowded field.  
Figure \ref{fighstcoh2}b shows the resultant 2.12\micron\ image and the spectroscopy slit.

We employed the NIRSPEC-6 blocking filter and selected
echelle and cross disperser positions to
detect several \h2\ rovibrational transitions,
\brg, and \ion{He}{1} $\lambda$2.06\micron\ in our spectroscopic observations.
We used the $0\farcs432\times24\arcsec$ slit at position angle 135\deg\ 
for resolution $\lambda/\Delta\lambda= 26,000$, where $\Delta\lambda$ is
the observed $FWHM$ of an unresolved emission line.
We obtained six exposures of 300 s each
on the night of 1999 April 30 (UT), nodding along the slit
between each exposure.  

We subtracted a median dark frame from each image, divided
by a flat field, and
interpolated over deviant pixels, removing both cosmic
rays and bad detector pixels.
The spectra required rectification in both the spectral and spatial
dimensions. 
We used the OH night sky lines and
wavenumbers tabulated by Abrams et al. (1994)\nocite{Abr94} for
wavelength calibration and the continuum emission of the telluric standard
for spatial rectification.
We observed HD 161056 and reduced the spectra similarly,
then modelled this B1.5V star 
as a $T=24,000$ K blackbody and interpolated over the 
stellar \brg\ feature.  We divided the M16 data by this
spectrum to correct for atmospheric absorption and
normalized the stellar spectrum from this $K=5.4$ standard at 2.2\micron\ 
to flux calibrate the data.

Figure \ref{figpvplots} contains the resultant two-dimensional spectra of several bright
lines.  We have subtracted off-source emission from each of these spectra.
For the molecular emission, the background is near the edge of the slit
toward the ionizing source, while for \brg, 
the background lies away from the ionizing source.
We list the complete set of detected lines, their vacuum wavelengths,
their total fluxes in the slit, and their peak fluxes in 
Table \ref{tablines}.  

\section{Physical Conditions\label{secphys}}
\subsection{Geometry\label{subsec:geom}}
The gross structure of the photodissociation region is evident in the
images (Fig. \ref{fighstcoh2}) and
two-dimensional spectra (Fig. \ref{figpvplots}).  
Fewer stars are observed directly through the cloud 
than in the surrounding region (Fig. \ref{fighstcoh2}b),
implying  $A_K > 1$.
From the center of the molecular cloud out to the photoionizing source,
distinct regions of quiescent molecular, excited molecular,
atomic, and ionized material appear.
The FUV photons
do not fully penetrate the molecular cloud, so the excited
\h2\ resides only on the cloud's surface.  
The optically-thin ionized emission
is brightest in the regions of highest density, near the cloud, and it is
spatially more extended than the \h2\ emission.
The slit is aligned toward the ionizing source and covers several edges
on the irregular surface of the cloud.  
Thus, we observe several distinct regions of photoevaporative flow. 
The greatest intensities
are detected in the regions that are viewed edge-on.

In these observations, each of the bright regions within the PDR consists of 
spatially segregated molecular and recombination line emission.
The two brightest regions at slit positions $-1\farcs2$ and $-3\farcs2$
in the 1-0S(1) emission, for example, are physically associated 
with the ionized emission at $7\farcs1$ and $3\farcs6$,
respectively (Fig. \ref{figpvplots}d).  
(Note, however, that the more spatially extended \brg\
peak at the top of slit includes some contribution from a nearby surface north
of the slit, which we do not observe in \h2.)
Moderate emission peaks appear at
$2\farcs8$ and $1\farcs8$, and fainter surfaces at 
$-9\farcs3$, $-7\farcs9$, and $5\farcs3$ in the 1-0S(1) line.  The first two of
these are blended in the spatially extended \brg\ emission, while
the latter correspond to local emission maxima at $-2\farcs8$, $-0\farcs2$,
and $11\farcs3$, so
the projected separations of the \h2\ and \brg\ emission
regions in the plane of the sky range from  
1.8 to $2.5\times10^{17}$ cm
at the 2000 pc distance of M16 (\cite{Hum78}; \cite{Hil93}).
The transition from the atomic to molecular zone typically occurs at $A_V\sim 2$
(\cite{Tie85}),
or in terms of column density, $N_H \sim 4\times 10^{21}{\rm\, cm^{-2}}$.
Thus, the average gas density in the bright regions of the PDR is
1.6--$2.2\times10^4{\rm\,cm^{-3}}$.

We determine the density in the ionized region and use
the molecular line width (discussed below) to calculate 
the average density in the molecular region of the PDR.
Bertoldi \& Draine (1996)\nocite{Ber96} show
$n_i=S_{Ly}/(4\pi R^2qc_i)$, where $S_{Ly}$ is the Lyman
continuum flux, $R$ is the distance to the ionizing source,
$q$ accounts for attenuation
of the ionization in the evaporative flow,
and $c_i$ is the isothermal sound speed in the ionized medium.
We determine the value of  
$q\simeq 4.0\times10^{-14}\sqrt{S_{Ly}r}/R$ (\cite{Ber89}),
where $r$ is the cloud radius.  
We find the density of the ionized region
$n_i=5.2\times10^3{\rm\,cm^{-3}}$, for $c_i=12\kms$ (calculated
for $T_{H{\sc ii}}=9500$ K, below) 
and $r=25\arcsec$.  
For pressure
balance across the ionization front, $n\simeq 2n_ic_i^2/(\Delta v/2.35)^2$,
where $\Delta v$ is the $FWHM$ of the molecular line width,
so in the molecular region, 
$n\simeq 3.2\times10^5\pcc$.
This is significantly greater than the density calculated above, which
integrated over contributions from the ionized
to the molecular region.  Here, $n$ is determined in the
molecular zone alone, which should be denser.

\centerline{\psfig{figure=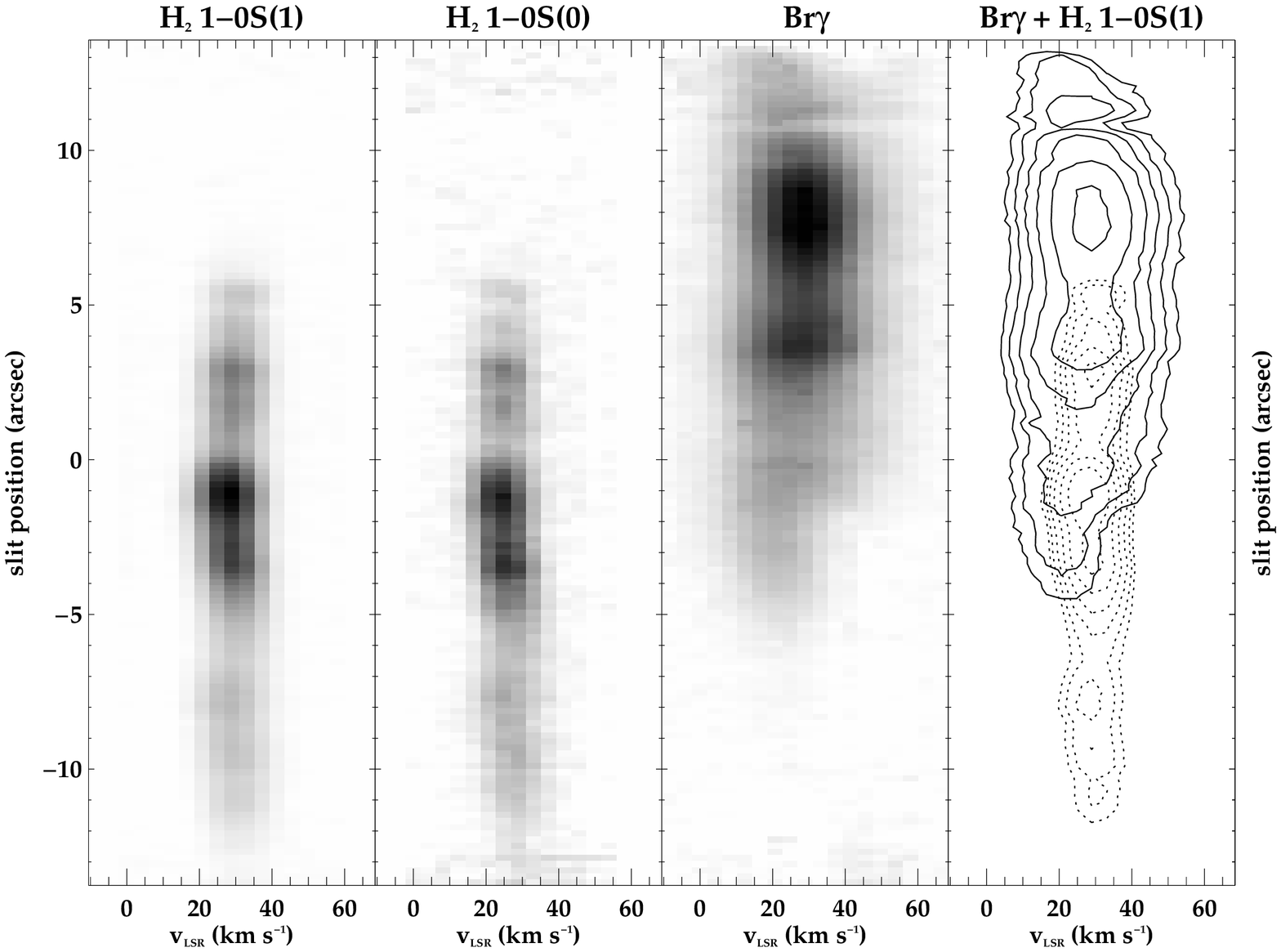,angle=90,width=3in}}
\figcaption{Two-dimensional spectra of bright lines.  
(a) \h2\ 1-0S(1), 2.12\micron, scaled linearly from 
0 to $5.0\times10^{-15}{\rm\,erg \,s^{-1}\,cm^{-2}\,arcsec^{-2}}$. 
(b)\h2\ 1-0S(0), 2.22\micron, scaled linearly from 
0 to $1.8\times10^{-15}{\rm\,erg \,s^{-1}\,cm^{-2}\,arcsec^{-2}}$. 
(c) \brg, 2.17\micron, scaled linearly from 
0 to $3.6\times10^{-15}{\rm\,erg \,s^{-1}\,cm^{-2}\,arcsec^{-2}}$. 
(d) Contours of \brg\ ({\it solid lines}) and \h2\ 1-0S(1) 
({\it dotted lines}) emission, in logarithmic intervals from
0.6 to $3.4\times10^{-15} {\rm\,erg \,s^{-1}\,cm^{-2}\,arcsec^{-2}}$.   
The \brg\ emission is broader and closer to the ionizing source 
(located at the top of the slit) than the molecular lines.
The two brightest regions in \brg\ at slit positions 
$3\farcs6$ and $7\farcs1$
correspond to the \h2\ peaks at
$-3\farcs2$ and $-1\farcs2$, respectively.
\label{figpvplots}}

\subsection{Kinematics}
The various gas phases exhibit distinct kinematic structure.
We measure the line profiles at slit position $3\farcs2$ in
1-0S(1) emission, which is clearly distinguished in the \brg\ 
and He emission at
$3\farcs6$.
Because we observe this PDR nearly edge-on, the central velocities
of the H$_2^*$ and ionized material are not significantly different; the
average motion occurs in the plane of the sky.  We measure
$v_{LSR}=29\kms$ for this 1-0S(1) feature and
$v_{LSR}=28\kms$ in \brg.
In \h2, $\Delta v =4.6\pm0.1\kms$, 
$\Delta v = 29.0\pm0.2\kms$ in \brg, and 
$\Delta v = 18.7\pm0.2\kms$ in \ion{He}{1} $\lambda 2.06\micron$,
after correcting for the instrumental profile.
In the molecular gas, the velocity width corresponds to a kinetic
temperature $T_{H_2}=930\pm 50$ K.  This value is reasonable near the edge
of the cloud in the H$_2^*$ zone where maximum temperature
of the PDR occurs (\cite{Tie85}).

Accounting for the mass difference between H and He 
demonstrates that these line widths are not purely thermal, however.
Assuming that they each consist of a macroscopic component that is 
independent of molecular weight in addition to the thermal component, we
find $\Delta v_{macro}=12.2 \kms$.  For \brg, $\Delta v_{therm}=25.6 \kms$,
and $T_{H{\sc ii}}=14000\pm230$ K.
This temperature is hotter than typically observed in \ion{H}{2}\ regions,
but several factors tend to increase $T_{H{\sc ii}}$ at the 
PDR-\ion{H}{2} interface:
increased photoelectric heating near a greater population of
grains;
the harder radiation field, which reduces the contribution of
significant low-ionization coolants, such as O$^+$;
and increased density, which reduces the
radiative cooling rate by increasing the collisional de-excitation
rate (e.g., \cite{Spi78}). 

The non-thermal recombination line widths may also be attributable to bulk
motion of material flowing off the cloud surface.  
In the bright region discussed above, 
we view the PDR nearly edge-on, along a line of
sight that includes both redshifted and blueshifted velocity components
of the material that evaporates off the curved surface of the cloud.
When the view of the PDR is closer to face-on, the intensity is
less enhanced, and only the blueshifted velocity component arises.
This is the case for the minor emission peaks
at the bottom of the slit, which occur on the near surface of 
the cloud.  These central velocities are slightly lower
(by approximately 7\kms) than those of
the brightest peaks, and the observed widths are narrower.
At slit position $-2\farcs8$, for example, we find
$\Delta v = 20.9\pm0.2$ and $10.1\pm0.2 \kms$ in \brg\ and \ion{He}{1}, 
respectively.
These both correspond to kinetic temperature $T_{H{\sc ii}}=9500\pm180$ K.
If this is a more accurate measurement of the temperature at 
the edge of the \ion{H}{2} region, then we have underestimated the
contribution of the non-thermal velocity component above.  While 
$T_{H{\sc ii}}$ at the interface is likely higher than
the typical temperatures in the diffuse central portions of \ion{H}{2}
regions, as argued above, it may be less than 14,000 K.

\subsection{H$_2$ Excitation}
In a PDR, UV pumping is the primary source of \h2\ excitation.  
In M16, such fluorescent emission obviously occurs.
A variety of lines that originate in high vibrational levels $v$
are observed, including 7-5O(4) and 3-2S(5) (Table \ref{tablines}), 
whereas thermal population of these high-$v$ levels
would dissociate the molecules.
We detect emission from all of the strong transitions
in our bandpass that
Black and van Dishoeck (1987)\nocite{Bla87} predict,
except those originating in levels 
$v \ge 9$ or lie in regions of poor atmospheric transparency.

Collisional excitation is also important,
however.  
The ratio of line intensities 
$I$(1-0S(1))/$I$(2-1S(1))$ = 2$--5, and the average ratio over the slit is 4.
As we found above in \S\ref{subsec:geom},
this requires $n>10^4\pcc$ (\cite{Bur90}), 
so collisions 
aid in populating the $v=1$ level.
We do not, however, 
observe line ratios as high as Allen et al. (1999)\nocite{All99},
who find $I$(1-0S(1))/$I$(2-1S(1))$ > 10$
in similar regions with Fabry-Perot imaging.  We expect our results
to be more accurate, having better flux calibration and simultaneous sky
measurements.

Line ratios also directly measure excitation temperatures.  
Across the slit, we find from the 1-0S(1) and 2-1S(1) lines 
vibrational temperature 
$T_{vib} = 3000$--3800 K, with a median $T_{vib} = 3400$ K.  
Using $I$(1-0S(1))/$I$(1-0S(3)), we find 
rotational temperature 
$T_{rot}= 1600$--2100 K and median $T_{rot}= 1800$ K.
Comparing $I$(1-0S(1))/$I$(1-0S(0)),  
$T_{rot}= 450$--870 K, with a median $T_{rot}= 580$ K.
The relatively low value of $T_{vib}$ and relatively
high value of $T_{rot}$ (particularly in the brightest
regions) distinguish the M16 PDR from 
purely fluorescent emission in a low density medium,
where $T_{vib}\sim 5500$ and $T_{rot}\sim 1400$ from
$I$(1-0S(1))/$I$(1-0S(3)) are typical.

We compare the line intensities of the three brightest regions with
the models of Draine \& Bertoldi (1996)\nocite{Dra96}.  
Considering all observed lines, all three regions are best fit with $\chi/n=0.1 {\rm\,cm^3}$,
and $T_0=1000$ K, where $\chi$ measures the incident UV flux ($\chi=1$ is the
Habing [1968]\nocite{Hab68} flux), and $T_0$ 
is the temperature at the edge of the PDR. 
The observed line ratios clearly rule
out models of a weak field ($\chi \le 10^3$) and low density ($n\le 10^4\pcc$).
Both $\chi= 10^4$ and $\chi=10^5$ are acceptable.
For consistency with other density measurements and the UV field expected
from members of NGC 6611, $\chi=10^4$ and $n=10^5\pcc$ 
is the preferred model.
Although the grid of models is coarse, we identify trends by contrasting these
regions.  
For example, the variation of density-sensitive line ratios along the spatial
extent of the slit demonstrates that the cloud density is not constant.
The decreased ratios
$I$(1-0S(1))/$I$(1-0S(0)) and $I$(1-0S(1))/$I$(2-1S(1)) at slit position
$-3\farcs2$ relative to slit position $-1\farcs2$ show that
the former region has lower density.

In thermal equilibrium, the ratio of ortho-hydrogen (having
odd rotational level, $J$)
to para-\h2\ (having even $J$) is 3, and only collisions 
can change the ortho:para ratio.
This abundance ratio is therefore a significant parameter in PDRs and affects
the emission line ratios. 
We calculate the  ratio of ortho-\h2\ and para-\h2\ 
from the 1-0S(1) and 1-0S(0) lines, 
finding ratios of 1.3--2.2, 
with an median value of 1.7.
Because UV pumping of the damping wings of optically thick lines
populate the excited levels, 
the expected ratio of column densities in the vibrationally
excited levels is 1.7, while the true ortho:para abundance is 3 (\cite{Ste99}).

The photoelectric effect on interstellar grains is likely to be the
primary heating source in PDRs, but its exact process is uncertain.
We determine the efficiency of converting FUV flux to photoelectric heating,
$\epsilon$, assuming thermal equilibrium.
Emission in the 
[\ion{O}{1}] 63\micron\ and [\ion{C}{2}] 158\micron\ fine structure lines
cools the gas.  For the measured temperature and density, we calculate a
net cooling rate 
$\Lambda = 4.2\times 10^{-17}{\rm\, erg\, s^{-1}\, cm^{-3}}$ in M16,
with 96\% due to [\ion{O}{1}].
In the parameterization of Bakes \& Tielens (1994)\nocite{Bak94},
the heating function 
$\Gamma = 10^{-24}n\epsilon\chi {\rm\, erg\, s^{-1}\, cm^{-3}}$.
Thus,  we find $\epsilon=0.042$, for $\chi=9700$ (\cite{All99}).
The theoretical prediction, $\epsilon=0.016$ (\cite{Bak94}),
is a function of $\chi\sqrt{T}/n_e$, where $n_e$
is the electron density.  We have assumed that the ionization of
carbon provides the electrons;  
increasing $n_e$ would increase $\epsilon$,
without significantly altering the cooling rate.
Alternatively, decreasing the gas to dust ratio
would allow a lower efficiency in equilibrium
with the cooling rate we predict.

\section{Conclusions\label{secconcl}}
Spectroscopy of M16 with NIRSPEC offers the advantage of simultaneous 
observation of both the molecular and ionized phases 
of its photodissociation regions.  At high resolution, we detect the spatial
and kinematic signatures of photoexcitation of molecular material.
Most of the emission is due to distinct regions of photodissociation 
on the molecular cloud's surface that we observe edge-on.
The spectrum includes several lines that originate in high vibrational
levels, which demonstrates that fluorescent excitation 
predominates. The density $n\sim10^5\pcc$ and varies among the
emitting regions, so collisions also preferentially populate
low-lying vibrational levels.
We measure kinetic temperature $T_{H_2}=930$ K in the molecular gas,
which requires
efficient photoelectric heating near the cloud's surface.
In the spatially-segregated ionized region, $T_{H{\sc ii}}=9500$ K,
and the velocity widths of the brightest regions 
include a macroscopic contribution.

\acknowledgements

It is a pleasure to acknowledge the hard work of past and present members
of the NIRSPEC instrument team at UCLA: Maryanne Angliongto, Oddvar
Bendiksen, George Brims, Leah Buchholz, John Canfield, Kim Chin, Jonah
Hare, Fred Lacayanga, Samuel B. Larson, Tim Liu, Nick Magnone, Gunnar
Skulason, Michael Spencer, Jason Weiss, and Woon Wong. In addition, we
thank the Keck Director Fred Chaffee, CARA instrument specialist Thomas A.
Bida, and all the CARA staff involved in the commissioning and integration
of NIRSPEC. We especially thank our Observing Assistants Joel Aycock, Gary
Puniwai, Charles Sorenson, Ron Quick, and Wayne Wack for their support.
We thank Amiel Sternberg
for helpful discussions and Marc Pound for providing his CO data.

\begin{deluxetable}{rllc}
\footnotesize
\tablewidth{0pt}
\tablecaption{Observed Lines\label{tablines}}
\tablehead{
\colhead{Transition}&\colhead{Wavelength}&\colhead{Total Flux}
&\colhead{Peak Surface Brightness} \cr 
&\multicolumn{1}{c}{[$\mu$m]} & \colhead{[${\rm erg\, s^{-1} cm^{-2}}$]}& 
\colhead{[${\rm erg\, s^{-1} cm^{-2} arcsec^{-2}}$]}}
\startdata
\multicolumn{1}{l}{H$_2$}\nl
7-5O(4)  &  1.94343  &  $3.9\pm0.2\times10^{-16}$  &  $1.9\pm0.4\times10^{-16}$  \nl 
2-1S(5)  &  1.94486  &  $1.1\pm0.01\times10^{-14}$  &  $3.2\pm0.2\times10^{-15}$  \nl
1-0S(3)\tablenotemark{a}  &  1.95756  &  $5.6\pm0.02\times10^{-14}$  &  $4.8\pm0.2\times10^{-15}$  \nl 
3-2S(5)  &  2.06556  &  $4.5 \pm0.06 \times 10^{-15}$  &  $3.7\pm0.6\times10^{-16}$  \nl 
8-6O(4)  &  2.12155  &  $7.4\pm0.2\times10^{-16}$  &  $1.2\pm0.3\times10^{-16}$  \nl 
1-0S(1)  &  2.12183  &  $6.2\pm0.02\times10^{-14}$  & $5.3\pm0.2\times10^{-15}$  \nl 
3-2S(4)  &  2.12799  &  $1.9\pm0.04\times10^{-15}$  & $2.4\pm0.5\times10^{-16}$  \nl 
1-0S(0)  &  2.22329  &  $2.4\pm0.01\times10^{-14}$  & $1.6\pm0.1\times10^{-15}$  \nl 
2-1S(1)  &  2.24772  &  $1.6\pm0.01\times10^{-14}$  & $1.3\pm0.1\times10^{-15}$  \nl 
\multicolumn{2}{l}{Recombination Lines}\nl
Br$\gamma$  &  2.16609  &  $1.3\pm0.003\times10^{-13}$  &  $3.7\pm0.2\times10^{-15}$  \nl 
Br$\delta$\tablenotemark{a,b}  &  1.94509  &  $2.3^{+1}_{-0.02}\times10^{-14}$  &  $1.3\pm0.1\times10^{-15}$  \nl 
\ion{He}{1}  &  2.05869  &  $9.7\pm0.03\times10^{-14}$  & $3.9\pm0.2\times10^{-15}$  \nl 
\ion{He}{1}  &  2.11259  &  $2.6\pm0.05\times10^{-15}$  & $1.9\pm0.4\times10^{-16}$  \nl 
\enddata
\tablenotetext{a}{Correction for atmospheric transparency is large, around 2}
\tablenotetext{b}{Missing flux due to poor atmospheric transparency}
\end{deluxetable}

\clearpage

\end{document}